\documentclass[prd,preprintnumbers,floatfix,nofootinbib,notitlepage]{revtex4}
\usepackage{amsfonts}
\usepackage{amssymb}
\usepackage{amsmath}
\usepackage{amsthm}

\begin{document}

\title{\bf On the covariance of teleparallel gravity theories}

\author{Alexey Golovnev${}^{1,2}$,  Tomi Koivisto${}^{3}$, Marit Sandstad${}^{3}$\\
{\small ${}^{1}${\it Faculty of Physics, St. Petersburg State University,}}\\ 
{\small\it Ulyanovskaya ul., d. 1, Saint Petersburg 198504, Russia,}\\
{\small ${}^{2}${\it St Petersburg National Research University of Information Technologies, Mechanics and Optics,}}\\ 
{\small\it Kronverkskiy pr. 49, Saint Petersburg, 197101, Russia,}\\
{\small agolovnev@yandex.ru}\\
{\small ${}^{3}${\it Nordita, KTH Royal Institute of Technology and Stockholm University,}}\\
{\small\it Roslagstullsbacken 23, SE-10691 Stockholm, Sweden,}\\
{\small tomik@astro.uio.no \qquad marit.sandstad@astro.uio.no}
}

\begin{abstract}

The basics of teleparallel gravity and its extensions are reviewed with particular emphasis on the problem of Lorentz-breaking choice of connection in 
pure-tetrad versions of the theories. Various possible ways to covariantise such models are discussed. A by-product is a new form of $f(T)$ field equations.

\end{abstract}

\preprint{NORDITA-2017-11}

\maketitle

\section{Introduction}

Well known issues with theoretical understanding of the gravitational interaction, ranging from hard problems of quantum gravity to overwhelmingly dark contents of the Universe, motivates us to seek a viable modified gravity model \cite{Clifton:2011jh}. Apparently a viable model should be theoretically meaningful, comply with existing experimental constraints, and ideally help us with some problems in quantum gravity and/or cosmology. However, even the first condition is often an obstacle. It is not so easy to modify gravity without inducing ghosts or other severe drawbacks.

It is natural in such a situation to try starting from another, though equivalent, formulation of General Relativity (GR). One option of this sort is actually known from the very early days. Exploiting the concept of absolute parallelism, as did Einstein\footnote{In an attempt at a unified theory, 
10 components of the tetrad were to represent the gravitational metric, and 
the remaining 6 were to be associated with electromagnetism. However, the latter 6 components are pure gauge, due to the inherent Lorentz invariance of the teleparallel structure. It comes without curvature, hence the name. The geometric foundations had been laid by Cartan \cite{Goenner2004}.}
 himself, one may describe gravity equivalently in terms of torsion instead
of the curvature, with the underlying field being the tetrad instead of the metric. This approach is known as teleparallel gravity. One of its advantages is that it allows to define the gravitational energy momentum tensor \cite{moller} Lorentz-covariantly \cite{newtensor}. In general, as the tetrad is intimately related to the gauge potential of translations, the field strength of which is the torsion, teleparallelism is a natural way to possible understanding of the gauge nature of gravity  \cite{blagojevic,Aldrovandi:2013wha}. 
  
A covariant metric action must contain second derivatives, but since this is not the case for a teleparallel tetrad action, it has been thought to be better suited for modifications because going, for example, to non-linear functions of the Lagragian density \cite{Linder:2010py} should not induce new, potentially pathological degrees of freedom. However, it turns out not to be that simple \cite{LSB1,LSB2,LSB3}. Setting the connection to vanish in the pure-tetrad formulation has broken the local Lorentz invariance. The effects add up to a mere surface term in the teleparallel equivalent of GR, but this fails to be the case in non-linear generalisations. Broken Lorentz invariance strikes back in the form of new degrees of freedom. Other issues have been encountered and discussed in for example \cite{Ferraro:2011us,Tamanini:2012hg,Ong:2013qja,Chen:2014qtl}.

The observation of the inviability of the pure-tetrad teleparallelism is not new at all (see e.g. \cite{kop}),
nor the fact that ''all the problematic features of the teleparallel theory become irrelevant by turning, as one should, to the full Poincar\'e gauge theory'' \cite{Blagojevic:2013xpa}. Nevertheless, while teleparallel modifications of gravity have become, especially due to the dark energy problem in cosmology, popular in recent years, we don't, however, find the paramount issue of the Lorentz invariance yet quite clarified in this context \cite{f(T)review}.
In recent discussions it has been pointed out that one can effectively restore the local Lorentz invariance by 
inserting an appropriate spin connection into the action since it will only introduce a surface term \cite{Krssak:2015rqa,Krssak:2015lba,Krssak}. An important prerequisite is that the spin connection has to be flat, or purely inertial. This is basically what one should understand as teleparallel gravity in the first place. It is covariant by construction, though not manifestly so in the gauge where the spin connection is set to zero. Any other gauge choice introduces a non-zero spin connection due to the inhomogeneous transformation: this is what is missed in the pure-tetrad formulation, where only the tetrad is transformed.


In this paper we aim at clarifying these points. In particular, we will focus on the technical problem of the variational formulation. In Sections II and III we give a brief but hopefully pedagogical introduction to teleparallel gravity. In Section IV we consider various covariantisations of the (pure-tetrad) teleparallel models and discuss extensions in Section V. In Section VI we also treat the case of $f(T)$ explicitly at the level of equations of motion. In Section we VII conclude with a summary.

\section{Tetrads and connections}

In the tetrad formulation, the metric at each point is associated with the set of tangent vectors via
\begin{equation}
\label{metric}
g_{\mu\nu}=e^a_{\mu}e^b_{\nu}\eta_{ab}
\end{equation}
which defines the tetrad fields $e^a_{\mu}$ up to local Lorentz rotations. Since we are interested in non-degenerate metrics, we assume that $e^a_{\mu}$ form a non-degenerate matrix, and inverse tetrads $e^{\mu}_{a}$ can be defined as the inverse matrix so that $e^a_{\mu}e^{\mu}_{b}\equiv\delta^a_b$ and $e^a_{\mu}e^{\nu}_{a}\equiv\delta^{\nu}_{\mu}$ with
$$g^{\mu\nu}=e_a^{\mu}e_b^{\nu}\eta^{ab}$$
for the inverse metric. 

The inverse tetrad can be viewed as a (reference) frame field. It represents a basis of tangent vectors at a given point of the spacetime manifold. It is a basis in which the quadratic form $g_{\mu\nu}$ acquires its canonical form $\eta_{ab}$ which is easily seen from $g_{\mu\nu}e^{\mu}_a e^{\nu}_b=\eta_{ab}$. Since such a basis always exists, it guarantees the tetrad representation (\ref{metric}) can always be found, locally. However, there may exist topological obstructions to global existence of a frame field. For instance, there are no smooth vector fields on a sphere, let alone a full set of those. Just like one cannot nicely comb a hairy ball, a tetrad description might not be globally possible for manifolds with non-trivial topology.

Now we can, if we want, consider every tensor with Latin indices instead of spacetime ones with the relation between the two being understood as
\begin{equation}
\label{getrans}
{\mathcal T}^{a_1,\ldots,a_n}_{b_1,\ldots,b_m}\equiv e^{a_1}_{\alpha_1}\cdots e^{a_n}_{\alpha_n}{\mathcal T}^{\alpha_1,\ldots,\alpha_n}_{\beta_1,\ldots,\beta_m} e^{\beta_1}_{b_1}\cdots e^{\beta_m}_{b_m}.
\end{equation}
One can also say that we have a copy of the tangent space at each point with the canonical metric $\eta_{ab}$ in it, and the tetrads realise an isomorphism between the two (pseudo)normed linear spaces via $A^{\mu}\to A^a=e^a_{\mu}A^{\mu}$.

Moreover, we can now have two types of the connection coefficients, $\Gamma^{\alpha}_{\mu\beta}$ for the usual tensors and $\omega^a_{\hphantom{a}\mu b}$ for those with tangent space indices. The latter of course must transform as a connection, whilst the tetrad by construction transforms as a tensor. 
Therefore, under a Lorentz rotation $\Lambda$,
\begin{equation}
\label{omtransL}
e^a_\mu  \longrightarrow \Lambda^a_c e^c_\mu\,, \quad
\omega^a_{\hphantom{a}\mu b} \longrightarrow \Lambda^a_c \omega^c_{\hphantom{a}\mu d}(\Lambda^{-1})^d_b-(\Lambda^{-1})^a_c\partial_{\mu}\Lambda^c_b\,.
\end{equation}
In order to freely change the nature of the indices by the tetrads, we wish this procedure to commute with taking a covariant derivative. Obviously, this goal would be achieved by the following requirement
\begin{equation}
\label{covconst}
\partial_{\mu}e^a_{\nu}+\omega^a_{\hphantom{a}\mu b}e^b_{\nu}-\Gamma^{\alpha}_{\mu\nu}e^a_{\alpha}=0
\end{equation}
which can be referred to as vanishing of the "full covariant derivative" of the tetrad. With this understanding in mind, we can conveniently use tensors with indices of both types, and the covariant derivatives would be unambiguously defined for a tensor even if we are allowed to transform from one type to another. The recipe is that we use $\Gamma$-terms for Greek indices, and $\omega$-terms for Latin indices: $\bigtriangledown_{\mu}T^{a\alpha}=\partial_{\mu}T^{a\alpha}+\Gamma^{\alpha}_{\mu\beta}T^{a\beta}+\omega^a_{\hphantom{a}\mu b}T^{b\alpha}$.

Condition (\ref{covconst}) is solved straightforwardly to obtain
\begin{equation}
\label{Gamma}
\Gamma^{\alpha}_{\mu\nu}=e_a^{\alpha}\left(\partial_{\mu}e^a_{\nu}+\omega^a_{\hphantom{a}\mu b}e^b_{\nu}\right)\equiv e_a^{\alpha}\ {\mathop\mathfrak D}_{\mu}e^a_{\nu}
\end{equation}
with $ {\mathop\mathfrak D}_{\mu}$ being the Lorentz-covariant (with respect to the Latin index only) derivative. Relation (\ref{Gamma}) can also be reversed to obtain the spin connection
$$\omega^a_{\hphantom{a}\mu b}=e^a_{\alpha}\Gamma^{\alpha}_{\mu\nu}e^{\nu}_b-e^{\nu}_b\partial_{\mu}e^a_{\nu}$$
which corresponds to a given affine connection on the manifold. In particular, one can find the spin connection $\mathop\omega\limits^{(0)}$ which corresponds to the Levi-Civita connection $\mathop\Gamma\limits^{(0)}(g)$ of a given metric $g$.

Basically, if (\ref{Gamma}) is valid, then both $\Gamma^{\alpha}_{\mu\beta}$ and $\omega^a_{\hphantom{a}\mu b}$ represent one and the same connection in different disguises. This conclusion is further substantiated by comparing the curvatures for both connections,
\begin{equation}
\label{spinRiemann}
R^a_{\hphantom{a}b\mu\nu}(\omega)=\partial_{\mu}\omega^a_{\hphantom{a}\nu b}-\partial_{\nu}\omega^a_{\hphantom{a}\mu b}+\omega^a_{\hphantom{a}\mu c}\omega^c_{\hphantom{c}\nu b}-\omega^a_{\hphantom{a}\nu c}\omega^c_{\hphantom{c}\mu b}
\end{equation}
and
\begin{equation}
\label{metricRiemann}
R^{\alpha}_{\hphantom{\alpha}\beta\mu\nu}(\Gamma)= \partial_{\mu}\Gamma^{\alpha}_{\nu\beta}-\partial_{\nu}\Gamma^{\alpha}_{\mu\beta} +\Gamma^{\alpha}_{\mu\rho}\Gamma^{\rho}_{\nu\beta}-\Gamma^{\alpha}_{\nu\rho}\Gamma^{\rho}_{\mu\beta}\,,
\end{equation}
which after a simple calculation gives
\begin{equation}
\label{corrRiemann}
R^{\alpha}_{\hphantom{a}\beta\mu\nu}(\Gamma)=e^{\alpha}_a R^a_{\hphantom{\alpha}b\mu\nu}(\omega) e^b_{\beta}\,.
\end{equation}
In other words, the two Riemann tensors are related by mere change of the types of indices. Therefore, those are one and the same tensor under our conventions which are common for all the tensors we use.

Note also that the non-metricity in this formalism (with the vanishing of the "full covariant derivative" of the tetrad) is automatically equal to zero because
\begin{equation}
\label{Qzero}
\bigtriangledown_{\alpha}g_{\mu\nu}=\eta_{ab}\left(\partial_{\alpha}\left(e^a_{\mu}e^b_{\nu}\right)-\Gamma^{\beta}_{\alpha\mu}e^a_{\beta}e^b_{\nu}-\Gamma^{\beta}_{\alpha\nu}e^a_{\mu}e^b_{\beta}\right)=-e^b_{\mu}e^c_{\nu}\left(\eta_{ab}\omega^a_{\hphantom{a}\alpha c}+\eta_{ac}\omega^a_{\hphantom{a}\alpha b}\right)=0
\end{equation}
where we have used (\ref{metric}), (\ref{covconst}), and the assumption that the matrices $\omega^{\cdot}_{\hphantom{a}\alpha \cdot}=\omega^a_{\hphantom{a}\alpha b}$ belong to the Lie algebra of the group $SO(1,3)$. 
A completely different approach is needed for incorporating symmetric teleparallel gravity \cite{NesterYo,Adak:2005cd}.

In particular, absence of non-metricity leads to antisymmetry of the Riemann tensor with respect to interchange of the first two indices, $R_{\alpha\beta\mu\nu}=-R_{\beta\alpha\mu\nu}$. In terms of the tetrad formulation (\ref{spinRiemann}) it follows from the antisymmetry of the spin connection coefficients with raised (or lowered) Latin indices, $\omega^{a\hphantom{\mu}b}_{\hphantom{a}\mu}=-\omega^{b\hphantom{\mu}a}_{\hphantom{b}\mu}$. And in the usual metric formalism the same conclusion can be achieved by noticing that from the basic definitions one derives the following identities:
\begin{eqnarray*}
\left[\bigtriangledown_{\mu}\ ,\ \bigtriangledown_{\nu}\right]A^{\alpha} & = &
R^{\alpha}_{\phantom{\alpha}\beta\mu\nu}A^{\beta}-T^{\beta}_{\hphantom{\beta}\mu\nu}\bigtriangledown_{\beta}A^{\alpha},\\
\left[\bigtriangledown_{\mu}\ ,\ \bigtriangledown_{\nu}\right]A_{\alpha}
& =& -R^{\beta}_{\phantom{\beta}\alpha\mu\nu}A_{\beta}-T^{\beta}_{\hphantom{\beta}\mu\nu}\bigtriangledown_{\beta}A_{\alpha}
\end{eqnarray*}
with $T^{\beta}_{\hphantom{\beta}\mu\nu}$ being the torsion tensor, see below. Now it is obvious that, if the metric $g_{\mu\nu}$ commutes with covariant derivatives, then the first two indices of the Riemann tensor enjoy the antisymmetry property even the in presence of torsion.

\section{Teleparallel vs pure-tetrad gravity}

In GR, the spin connection represents both gravitation and inertial effects. The equivalence principle guarantees the existence of a local frame,
where such a connection vanishes. 

In teleparallelism, gravitation is described in terms of torsion instead of curvature. The spin connection is then associated solely with 
inertia. This interpretation is naturally inherited from special relativity: considering the equation of motion for free a particle in the Minkowski space, $d u^a/ds=0$, one obtains, in a general class of frames, $d u^a/ds = \omega^a_{\phantom{a}\mu b}u^b u^\mu$, where $\omega^a_{\phantom{a}\mu b}$ is just the teleparallel spin connection.  
Accordingly, it is always (barring topological
complications) possible to choose a global class of frames where $\mathop\omega\limits^{\mathfrak W}{\vphantom{\omega}}^a_{\hphantom{a}\mu b}=0$. 
Obviously, in this frame the field strength of the connection (\ref{spinRiemann}), and the spacetime curvature (\ref{metricRiemann}), vanish. 
A Lorentz transformation (\ref{omtransL}) introduces a non-zero inertial connection\footnote{To further substantiate the interpretation as inertia, switch off gravity by setting the Newton's constant to zero. The anholonomy of the tetrads, $[e_a,e_b] = f^c_{\phantom{c}ab}e_c$, then determines 
the spin connection as $\omega^a_{\mu b} = \frac{1}{2}(f_{b\phantom{a}c}^{\phantom{b}a} + f_{c\phantom{a}b}^{\phantom{c}a} - f^a_{\phantom{a}bc})e^c_\mu$. 
This also shows that the spin connection of teleparallel gravity is not arbitrary. }, but no curvature.  
An equivalent description of general relativity is then given in terms of the torsion tensor
\begin{equation}
\label{torsion}
T^{\alpha}_{\hphantom{\alpha}\mu\nu}\equiv\Gamma^{\alpha}_{\mu\nu}-\Gamma^{\alpha}_{\nu\mu}
\end{equation}
instead of curvature. 
Below we outline the ingredients of this approach that are needed in the following study of the action formulation. For a more thorough introduction to teleparallel gravity, see \cite{Aldrovandi:2013wha}. 

\subsection{The formulation of teleparallel gravity}

Assuming that $\bigtriangledown_{\alpha}g_{\mu\nu}=0$, one can follow the standard textbook derivation of the Levi-Civita connection and prove that
\begin{equation}
\label{Gammvar}
\Gamma^{\alpha}_{\mu\nu}=\mathop\Gamma\limits^{(0)}{\vphantom{\Gamma}}^{\alpha}_{\mu\nu}(g)+K^{\alpha}_{\hphantom{\alpha}\mu\nu}
\end{equation}
where $\mathop\Gamma\limits^{(0)}{\vphantom{\Gamma}}^{\alpha}_{\mu\nu}(g)$ is the Levi-Civita connection of the metric $g$, while the tensor $K$
\begin{equation}
\label{contortion}
K_{\alpha\mu\nu}=\frac12\left(T_{\alpha\mu\nu}+T_{\nu\alpha\mu}+T_{\mu\alpha\nu}\right)=\frac12\left(T_{\mu\alpha\nu}+T_{\nu\alpha\mu}-T_{\alpha\nu\mu}\right),
\end{equation}
is known under the name of contortion. It is obviously antisymmetric with respect to two indices:
\begin{equation}
\label{Ksymm}
K_{\alpha\mu\nu}=-K_{\nu\mu\alpha}.
\end{equation}

Substituting connection (\ref{Gammvar}) into the the definition of curvature, we get
\begin{equation}
\label{Riemmvar}
R^{\alpha}_{\hphantom{\alpha}\beta\mu\nu}(\Gamma)=R^{\alpha}_{\hphantom{\alpha}\beta\mu\nu}(\mathop\Gamma\limits^{(0)})+ \mathop\bigtriangledown\limits^{(0)}{\vphantom{\bigtriangledown}}_{\mu}K^{\alpha}_{\hphantom{\alpha}\nu\beta}-\mathop\bigtriangledown\limits^{(0)}{\vphantom{\bigtriangledown}}_{\nu}K^{\alpha}_{\hphantom{\alpha}\mu\beta} +K^{\alpha}_{\hphantom{\alpha}\mu\rho}K^{\rho}_{\hphantom{\rho}\nu\beta}-K^{\alpha}_{\hphantom{\alpha}\nu\rho}K^{\rho}_{\hphantom{\rho}\mu\beta}
\end{equation}
for the Riemann tensor with  $ \mathop\bigtriangledown\limits^{(0)}{\vphantom{\bigtriangledown}}_{\mu}$ being the covariant derivative associated to $\mathop\Gamma\limits^{(0)}{\vphantom{\Gamma}}^{\alpha}_{\mu\nu}(g)$. Making the necessary contractions we obtain the scalar curvature
\begin{equation}
\label{telepid}
R(\Gamma)=R(\mathop\Gamma\limits^{(0)})+2 \mathop\bigtriangledown\limits^{(0)}{\vphantom{\bigtriangledown}}_{\mu}T^{\mu}+\mathbb T
\end{equation}
where the torsion vector is
\begin{equation}
\label{Tvector}
T_{\mu}\equiv T^{\alpha}_{\hphantom{\alpha}\mu\alpha}=-T^{\alpha}_{\hphantom{\alpha}\alpha\mu},
\end{equation}
and the torsion scalar can be written in several equivalent ways:
\begin{equation}
\label{Tscalar}
{\mathbb T} \equiv  \frac12 K_{\alpha\beta\mu}T^{\beta\alpha\mu}-T_{\mu}T^{\mu}
= \frac12 T_{\alpha\beta\mu}S^{\alpha\beta\mu}
=  \frac14 T_{\alpha\beta\mu}T^{\alpha\beta\mu}+\frac12 T_{\alpha\beta\mu}T^{\beta\alpha\mu}-T_{\mu}T^{\mu}
\end{equation}
with the superpotential
\begin{equation}
\label{superpot}
S^{\alpha\mu\nu}\equiv K^{\mu\alpha\nu}+g^{\alpha\mu}T^{\nu}-g^{\alpha\nu}T^{\mu}
\end{equation}
which satisfies the antisymmetry condition $S^{\alpha\mu\nu}=-S^{\alpha\nu\mu}$. One can also write it down in terms of three possible irreducible representations of the Lorentz group (vector, axial (fully antisymmetric) and pure tensor parts), all with non-zero coefficients \cite{Aldrovandi:2013wha} which shows, in particular, that ${\mathbb T}$ is a non-degenerate quadratic form of the torsion components.

As stated above, the teleparallel equivalence principle allows to exploit the Weitzenb{\" o}ck connection given by $\mathop\omega\limits^{\mathfrak W}{\vphantom{\omega}}^a_{\hphantom{a}\mu b}=0$ or
\begin{equation}
\label{GWeitz}
\mathop\Gamma\limits^{\mathfrak W}{\vphantom{Gamma}}^{\alpha}_{\mu\nu}= e^{\alpha}_a \partial_{\mu}e^a_{\nu}
\end{equation}
which is obviously curvature-free, $R^{\alpha}_{\hphantom{\alpha}\beta\mu\nu}(\mathop\Gamma\limits^{\mathfrak W})=0$. 
We can denote the determinant of $e^a_{\mu}$ by $\| e\|$, and see from (\ref{telepid}) that the action
\begin{equation}
\label{telepact}
S_{\mathfrak W}=-\int d^4 x \| e\|\cdot {\mathbb T}
\end{equation}
is equivalent to the action of GR, $\int d^4 x \sqrt{-g}\cdot R(\mathop\Gamma\limits^{(0)})$, modulo the surface term if the Weitzenb{\" o}ck, or any other inertial, connection is assumed. Moreover, if one forgets about the underlying geometry and views these actions as functionals of fields $e^a_{\mu}$, then those are simply equal (modulo surface terms). In particular, for the teleparallel equivalent of GR the local Lorentz invariance is restored even with the fixed connection at the level of the action up to the surface terms.\footnote{It is thus more precise to say that the teleparallel equivalent of GR is {\it quasi Lorentz invariant}, when the surface terms are taken into account. In principle the theory can be built to include a term that cancels this boundary term, as was discussed in the Ref. \cite{Bahamonde:2015hza}. There the approach was also extended to $f(T)$-theories. If the functional dependencies on the $T$-term and the boundary term are matched suitably, this yields Lorentz invariant versions of $f(T)$-theories.}

The historical version of the theory \cite{moller}, which is sometimes referred to as ''pure-tetrad (teleparallel) gravity'', can be  defined by absence of the spin connection. Note that this definition blatantly breaks local Lorentz invariance because the connection is not a tensor, and it's vanishing is not a 
covariant condition (whereas, the vanishing of the curvature is). More formally, if in a chosen frame the spin connection vanishes, then according to the formula (\ref{omtransL}), it must acquire the form of a purely inertial connection $\omega^a_{\hphantom{a}\mu b}=-(\Lambda^{-1})^a_c\partial_{\mu}\Lambda^c_b$ in other frames. In consistent teleparallel gravity, the connection (\ref{GWeitz}) is not a definition but a - physically motivated - gauge choice.  
The connection is always purely inertial and subject to the rule (\ref{omtransL}).

A big portion of current literature is written solely in terms of the Weitzenb{\" o}ck connection (\ref{GWeitz}). 
Often the pure-tetrad approach is assumed explicitly, often it is not clear what is assumed. 
The practical difference of the approaches is that the Lorentz transformation in the pure-tetrad version takes into account only the first 
part of (\ref{omtransL}), but leaves the spin connection trivial. As it should be clear at this point, this lacks conceptual support.
One has to regard a pure-tetrad model as valid description of physics in only a single, fixed frame. 
Bestowed an inertial connection, the model would become able to describe the same physics in any locally Lorentz rotated frame, 
with the spin connection coefficients then representing the inertial effects of choosing an anholonomic coordinate system.


\subsection{Equations of motion with fixed connection}

In this subsection, we derive the equations of motions for the pure-tetrad gravity.
We then have an arbitrary, though fixed, spin connection. In particular, 
we have the following first order variations for the inverse tetrad, measure, metric and torsion:
\begin{eqnarray}
\label{inve}
\delta e^{\mu}_a & = & -e^{\mu}_b e^{\nu}_a \delta e^b_{\nu},\\
\label{edetvar}
\delta\|e\| & = & \|e\|\cdot e^{\mu}_a \delta e^a_{\mu},\\
\label{emetvar}
\delta g_{\mu\nu} & = & \eta_{ab}\left( e^a_{\mu}\delta  e^b_{\nu}+ e^a_{\nu} \delta e^b_{\mu}\right),\\
\label{einvvar}
\delta g^{\mu\nu} & = & -\left(g^{\mu\alpha}e^{\nu}_{a}+g^{\nu\alpha}e^{\mu}_{a}\right)\delta e^a_{\alpha},\\
\label{etorvar}
\delta_e T^{\alpha}_{\hphantom{\alpha}\mu\nu} & = & -e^{\alpha}_a T^{\beta}_{\hphantom{\beta}\mu\nu}\delta e^a_{\beta}+e^{\alpha}_a \left({\mathop\mathfrak D}_{\mu}\delta e^a_{\nu}-{\mathop\mathfrak D}_{\nu}\delta e^a_{\mu}\right).
\end{eqnarray}
It is quite simple to find the first order variations of the following quadratic combinations of torsion components:
\begin{eqnarray}
\label{1evar}
\delta_e (T_{\mu}T^{\mu}) & =  & -2\left(T^{\beta}T^{\alpha}_{\hphantom{\alpha}\beta\mu}+T^{\alpha}T_{\mu}\right)e^{\mu}_a \delta e^a_{\alpha}+2\left(T^{\alpha}e^{\mu}_a-T^{\mu}e^{\alpha}_a \right)\cdot {\mathop\mathfrak D}_{\alpha}\delta e^a_{\mu}\\
\label{2evar}
\delta_e (T_{\alpha\mu\nu}T^{\mu\alpha\nu}) & = & 2\left(T^{\beta\mu\alpha}-T^{\alpha\mu\beta}\right)T_{\mu\alpha\nu} e^{\nu}_a \delta e^a_{\beta}+\left(T^{\alpha\hphantom{\mu}\beta}_{\hphantom{\alpha}\mu}-T^{\beta\hphantom{\mu}\alpha}_{\hphantom{\beta}\mu}\right)e^{\mu}_a \cdot{\mathop\mathfrak D}_{\alpha}\delta e^a_{\beta}\\
\label{3evar}
\delta_e (T_{\alpha\mu\nu}T^{\alpha\mu\nu}) & = & 4T_{\alpha}^{\hphantom{\alpha}\mu\nu}e^{\alpha}_a \cdot{\mathop\mathfrak D}_{\mu}\delta e^a_{\nu}-4T^{\alpha\mu\nu}T_{\alpha\mu\beta}e^{\beta}_a \delta e^a_{\nu}
\end{eqnarray}
These should be enough to derive equations of motion for any conceivable  model of interest.

In particular, for the teleparallel equivalent of GR (\ref{telepact}) we have
$$\delta_e S=-\int d^4 x \|e\|\cdot\left(-2S^{\alpha\mu\nu}T_{\alpha\beta\nu}e^{\beta}_a \delta e^a_{\mu}+{\mathbb T}e^{\mu}_a \delta e^a_{\mu} -2S_{\beta}^{\hphantom{\beta}\mu\alpha}e^{\beta}_a {\mathop\mathfrak D}_{\alpha}\delta e^a_{\mu}\right)$$
with the Lorentz-covariant derivative ${\mathop\mathfrak D}$ being equal to the ordinary one, since $\omega^b_{\hphantom{b}\alpha a}=0$ in the Weitzenb{\" o}ck case. We need to perform integration by parts in the last term which gives
$$2\delta e^a_{\mu}\cdot \left(\partial_{\alpha}\left(\|e\|\cdot S_{\beta}^{\hphantom{\beta}\mu\alpha}e^{\beta}_a\right)-\|e\|\cdot \omega^b_{\hphantom{b}\alpha a}S_{\beta}^{\hphantom{\beta}\mu\alpha}e^{\beta}_b\right)=2\|e\|\cdot\left(\mathop\bigtriangledown\limits^{(0)}{\vphantom{\bigtriangledown}}_{\alpha} S_{\beta}^{\hphantom{\beta}\mu\alpha}-K^{\nu}_{\hphantom{\nu}\alpha\beta}S_{\nu}^{\hphantom{\nu}\mu\alpha}\right)\cdot e^{\beta}_a\delta e^a_{\mu}$$
where we have used the antisymmetry of $S$ and corrected for the difference between $\Gamma$ and $\mathop\Gamma\limits^{(0)}$ by the second term on the right hand side. Indeed, due to the antisymmetry of $S$ we have
$$\mathop\bigtriangledown\limits^{(0)}{\vphantom{\bigtriangledown}}_{\nu} S_{a}^{\hphantom{a}\mu\nu}=\frac{1}{\|e\|}\partial_{\nu}\left(\|e\|S_a^{\mu\nu}\right)-\mathop\omega\limits^{(0)}{\vphantom{omega}}^b_{\hphantom{b}\nu a}S_{b}^{\hphantom{b}\mu\nu}$$
and correct for the different connection by noting that $\omega^b_{\hphantom{b}\nu a}-\mathop\omega\limits^{(0)}{\vphantom{omega}}^b_{\hphantom{b}\nu a}=K^b_{\hphantom{b}\nu a}$.

Finally, using the non-degeneracy of tetrads, we get the equations of motion in the form
\begin{equation}
\label{telepeom}
\mathop\bigtriangledown\limits^{(0)}{\vphantom{\bigtriangledown}}_{\alpha} S_{\beta}^{\hphantom{\beta}\mu\alpha}-S^{\alpha\mu\nu}\left(T_{\alpha\beta\nu}+K_{\alpha\nu\beta}\right)+\frac12 {\mathbb T}\delta^{\mu}_{\beta}=0
\end{equation}
which can be shown to be equivalent to general relativity by direct substitution of
$$R^{\alpha}_{\hphantom{\alpha}\beta\mu\nu}(\mathop\Gamma\limits^{(0)})=-\left( \mathop\bigtriangledown\limits^{(0)}{\vphantom{\bigtriangledown}}_{\mu}K^{\alpha}_{\hphantom{\alpha}\nu\beta}-\mathop\bigtriangledown\limits^{(0)}{\vphantom{\bigtriangledown}}_{\nu}K^{\alpha}_{\hphantom{\alpha}\mu\beta} +K^{\alpha}_{\hphantom{\alpha}\mu\rho}K^{\rho}_{\hphantom{\rho}\nu\beta}-K^{\alpha}_{\hphantom{\alpha}\nu\rho}K^{\rho}_{\hphantom{\rho}\mu\beta}\right)$$
into the Einstein equation,
$$G^{\mu}_{\beta}=0.$$

The coupling to matter can be prescribed consistently \cite{Obukhov:2002tm}. Bosonic fields interact with the metric (\ref{metric}), and variation with respect to tetrads gives the energy momentum tensor with one index renamed to Latin. However, the fermionic fields require spin connection, and the most reasonable way to do that is to use the $\mathop\omega\limits^{(0)}$ connection. In the pure-tetrad teleparallel gravity this seems somewhat ad hoc, but in a larger context of e.g. Poincar{\' e} theory, the prescribed connection can be viewed as the full (i.e. Cartan) connection whose difference from the vertical (i.e. Ehresmann) connection is indeed the contortion.

\section{Covariant teleparallel gravity}

As described above, the pure-tetrad gravity makes use of the Lorentz-breaking, a.k.a. Weitzenb{\" o}ck prescription (\ref{GWeitz}) for connection. It is relatively benign when dealing with the teleparallel equivalent of general relativity\footnote{A local Lorentz violation of this sort, is not catastrophic from an experimental viewpoint. Theoretically, it's very unpleasant indeed, and it goes against the very idea of tetrads which calls for the freedom of choice of the frames. However, it is not the same as violating the Lorentz invariance for the physical fields which correspond to elementary particles. If one considers a regime in which the gravitational interaction of particles is negligible and one can see that they move freely in a Minkowski space with zero curvature and zero torsion, then it is not observable whether it was necessary to choose some special fixed frame for tetrads or not.}, but strikes back in the form of new degrees of freedom with any attempts at generalisation like that of $f(T)$ \cite{f(T)review}. Of course, it can be made superficially covariant by admitting an arbitrary inertial spin connection into the action (\ref{telepact}), which is kept fixed and not varied while deriving the equations of motion. This is similar to hiding a fundamental anisotropy by explicitly introducing a fixed vector field in the direction of the anisotropy into the action.

However, a stronger statement can be made. As claimed in \cite{Krssak} and repeated in the review paper \cite{f(T)review}, the model can be Lorentz-covariantised by allowing for a spin-connection as a new variable because the spin-connection enters the teleparallel action only as a surface term. In other words, the variation with respect to the spin connection does not produce additional equations of motion. It is crucial that the spin connection here should be understood as a purely inertial one. Otherwise, as is easily seen from (\ref{telepid}), the scalar curvature of the spin connection pops up. In fact, arguments against a trivial covariantisation, without the constraint on inertiality, were already given in \cite{kop,LSB3,Krssak:2015rqa}.

In the rest of this Section we consider several approaches to variations with respect to the spin connection in actions of the teleparallel type. Though restricting for simplicity to the action corresponding to GR in some explicit examples, the conclusions are generalised in the following Section.

\subsection{Variation with respect to independent connection}

Variations with respect to the spin connection coefficients can be derived exactly since
\begin{equation}
\label{omtorvar}
\delta_{\omega} T^{\alpha}_{\hphantom{\alpha}\mu\nu}=\delta\omega^{\alpha}_{\hphantom{\alpha}\mu\nu}-\delta\omega^{\alpha}_{\hphantom{\alpha}\nu\mu},
\end{equation}
is an exact relation for $\delta\omega^{\alpha}_{\hphantom{\alpha}\mu\nu}\equiv e^{\alpha}_a e_{\nu}^b \delta\omega^{a\hphantom{\mu}}_{\hphantom{a}\mu b}$. We have
\begin{eqnarray}
\label{omtorvecvar}
\delta_{\omega}T_{\mu}=\delta_{\omega} T^{\alpha}_{\hphantom{\alpha}\mu\alpha} & = & -\delta\omega^{\alpha}_{\hphantom{\alpha}\alpha\mu},\\
\label{1omvar}
\delta_{\omega}\left(T^{\mu}T_{\mu}\right) & = & -2T_{\alpha}\delta\omega_{\mu}^{\hphantom{\mu}\mu\alpha}+\delta\omega_{\mu}^{\hphantom{\mu}\mu\alpha}\cdot\delta\omega^{\nu}_{\hphantom{\nu}\nu\alpha},\\
\label{2omvar}
\delta_{\omega}\left(T^{\alpha\mu\nu}T_{\mu\alpha\nu}\right) & = & 2\left(T_{\mu\alpha\nu}-T_{\alpha\mu\nu}\right)\delta\omega^{\alpha\mu\nu}-\left(\delta\omega_{\alpha\mu\nu}-3\delta\omega_{\mu\alpha\nu}\right)\cdot\delta\omega^{\alpha\mu\nu},\\
\label{3omvar}
\delta_{\omega}\left(T^{\alpha\mu\nu}T_{\alpha\mu\nu}\right) & = & 4T_{\alpha\mu\nu}\delta\omega^{\alpha\mu\nu}+2\left(\delta\omega_{\alpha\mu\nu}-\delta\omega_{\mu\alpha\nu}\right)\cdot\delta\omega^{\alpha\mu\nu}.
\end{eqnarray}
where we have used that, given the symmetry properties, there are only two independent contractions of $T$ and $\delta\omega$, namely $T_{\mu\alpha\nu}\delta\omega^{\alpha\mu\nu}$ and $T_{\alpha\mu\nu}\delta\omega^{\alpha\mu\nu}$, and similarly for contractions of two $\delta\omega$-s.

Suppose now that we want to covariantise the teleparallel action by allowing for an arbitrary spin connection in the torsion scalar,
\begin{equation}
\label{covart}
S=-\int d^4 x \| e\|\cdot {\mathbb T}(e,\omega),
\end{equation}
and varying independently with respect to both variables $e$ and $\omega$.

Using the variations (\ref{1omvar}) -- (\ref{3omvar}) and relation (\ref{Tscalar}), we have 
$$\delta_{\omega} S=-\int d^4 x \| e\|\cdot\left(T^{\mu}_{\hphantom{\mu}\alpha\nu}+2T_{\nu}\delta_{\alpha}^{\mu}\right)\delta\omega^{\alpha\hphantom{\mu}\nu}_{\hphantom{\alpha}\mu}.$$
The equation of motion is
$$T^{\mu}_{\hphantom{\mu}\alpha\nu}+T_{\nu}\delta_{\alpha}^{\mu}-T_{\alpha}\delta_{\nu}^{\mu}=0$$
which (in dimension $d\neq 2$) entails $T_{\mu}=0$ upon tracing, and totally
$$T^{\mu}_{\hphantom{\mu}\alpha\nu}=0.$$
Therefore this covariantisation procedure does not give a desired result.

\subsection{Teleparallel action with inertial spin connection}

A better idea would be to vary the spin connection in the inertial class only. The latter can be imposed by demanding
\begin{equation}
\label{inertial}
\omega^a_{\hphantom{a}\mu b}=-(\Lambda^{-1})^a_c \partial_{\mu}\Lambda^c_b
\end{equation}
 where $\Lambda$ is an arbitrary Lorentz matrix and varying
\begin{equation}
\label{covtelep}
S_{{\mathfrak W}^{\prime}}=-\int d^4 x \| e\|\cdot {\mathbb T}(e,\omega(\Lambda))
\end{equation}
with respect to $e$ and $\Lambda$. Literally it means that there exists a frame in which $\omega=0$ (Weitzenb{\" o}ck), however one is allowed to make a local Lorentz rotation by an arbitrary matrix field $\Lambda^a_b (x)$ whose values belong to Lorentz group and which produces connection coefficients given by (\ref{inertial}). Explicit calculations are given below. However, the essence is very simple. Varying the spin connection with tetrads fixed does not change the Levi-Civita connection, and from (\ref{telepid}) it follows that
$$\delta_{\Lambda}\mathbb T=\delta_{\Lambda} R(\omega)-2\mathop\bigtriangledown\limits^{(0)}{\vphantom{\bigtriangledown}}_{\mu}(\delta_{\Lambda} T^{\mu})$$
where $\delta_{\Lambda}(...)=\delta_{\omega}(...)\cdot \delta_{\Lambda}\omega$.
And since $R(\omega(\Lambda))\equiv 0$, variation $\delta_{\omega}S_{{\mathfrak W}^{\prime}}$ of the action (\ref{covtelep}) is a surface term and does not produce any new equation of motion. The model, though locally Lorentz covariant, is then equivalent to teleparallel gravity.

Let us give this statement some more details. The variation $ \delta_{\Lambda}\omega$ is performed by applying an arbitrary Lorentz transformation to $\Lambda$ which can be written as $\Lambda\to (\exp{\lambda})\cdot\Lambda$ and $\Lambda^{-1}\to\Lambda^{-1}\cdot\exp(-\lambda)$ where $\lambda^a_b$ is a (infinitesimal) matrix from the algebra of the Lorentz group\footnote{It is {\it not} the variation which corresponds to rotation $e\to (\exp{\lambda})\cdot e$. However, every two inertial connections are different by a choice of the matrix $\Lambda$, and since the Lorentz group is indeed a group one can convert one into another multiplying by a suitable group element.}, $\lambda^{ab}=-\lambda^{ba}$ with $\lambda^{ab}\equiv\eta^{bc}\lambda^a_c$. Then, at the first order,
\begin{equation}
\label{inertvar}
\delta\omega^a_{\hphantom{a}\mu b}=-(\Lambda^{-1})^a_c (\partial_{\mu}\lambda^c_d)\Lambda^d_b.
\end{equation}
In the Weitzenb{\" o}ck frame with $\Lambda=I$ we simply have $\delta\omega^a_{\hphantom{a}\mu b}=-\partial_{\mu}\lambda^a_b$. Otherwise one can also look at the variation (\ref{inertvar}) differently:
\begin{equation*}
-\delta\omega^a_{\hphantom{a}\mu b}=\partial_{\mu}\left((\Lambda^{-1})^a_c\lambda^c_d\Lambda^d_b\right)- (\partial_{\mu}(\Lambda^{-1})^a_c)\lambda^c_d\Lambda^d_b-(\Lambda^{-1})^a_c \lambda^c_d(\partial_{\mu}\Lambda^d_b)={\mathfrak D}_{\mu}{\tilde\lambda}^a_b
\end{equation*}
where ${\tilde\lambda}^a_b\equiv(\Lambda^{-1})^a_c\lambda^c_d\Lambda^d_b$ is the Lorentz  transformation of $\lambda$ to another frame, and $\mathfrak D$ is the Lorentz-covariant derivative with flat connection (\ref{inertial}).

Now, let us look at the variation of the 4-divergence of the torsion vector:
$$\delta_{\omega}\left(\|e\| \mathop\bigtriangledown\limits^{(0)}{\vphantom{\bigtriangledown}}_{\mu}T^{\mu}\right)=\delta_{\omega}\left(\vphantom{\int}\partial_{\mu}\left(\|e\|T^{\mu}\right)\right)=-\partial_{\mu}\left(\|e\|e^{\alpha}_a\delta\omega^{a\hphantom{\alpha}b}_{\hphantom{a}\alpha}e^{\mu}_b\right).$$
One can differentiate the right hand side, and using $\partial_{\mu}\|e\|=\|e\|\cdot e^{\beta}_c\partial_{\mu}e^c_{\beta}$, and $\partial_{\mu}e^{\alpha}_a=-e^{\beta}_a e^{\alpha}_c\partial_{\mu}e^c_{\beta}$ and analogously for $\partial_{\mu}e^{\mu}_b$, we get
$$-\partial_{\mu}\left(\|e\|\cdot e^{\alpha}_a\delta\omega^{a\hphantom{\alpha}b}_{\hphantom{a}\alpha}e^{\mu}_b\right)=\|e\|\cdot\left(\vphantom{\int}e^{\alpha}_c(\partial_{\mu}e^c_{\beta})\delta\omega^{\beta\hphantom{\alpha}\mu}_{\hphantom{\beta}\alpha}+(e^{\mu}_c\partial_{\mu}e^c_{\beta}-e^{\mu}_c\partial_{\beta}e^c_{\mu})\delta\omega^{\alpha\hphantom{\alpha}\beta}_{\hphantom{\alpha}\alpha}-e^{\alpha}_a e^{\mu}_b\partial_{\mu}\delta\omega^{a\hphantom{\alpha}b}_{\hphantom{a}\alpha}\right)$$
with $\delta\omega^{\beta\hphantom{\alpha}\mu}_{\hphantom{\beta}\alpha}\equiv e^{\beta}_a e^{\mu}_b\delta\omega^{a\hphantom{\alpha}b}_{\hphantom{a}\alpha}$. Given the antisymmetry of $\omega$ and the definition of the Weitzenb{\" o}ck torsion ${\mathop T\limits^{\mathfrak W}}{\vphantom{T}}^{\mu}_{\hphantom{\mu}\alpha\nu}=e^{\mu}_c\partial_{\alpha}e^c_{\nu}-e^{\mu}_c\partial_{\nu}e^c_{\alpha}$ we derive from this relation
\begin{equation}
\label{Identity}
\| e\|\cdot\left({\mathop T\limits^{\mathfrak W}}{\vphantom{T}}^{\mu}_{\hphantom{\mu}\alpha\nu}+2{\mathop T\limits^{\mathfrak W}}{\vphantom{T}}_{\nu}\delta_{\alpha}^{\mu}\right)\delta\omega^{\alpha\hphantom{\mu}\nu}_{\hphantom{\alpha}\mu}=2\partial_{\mu}\left(\|e\|\cdot e^{\alpha}_a\delta\omega^{a\hphantom{\alpha}b}_{\hphantom{a}\alpha}e^{\mu}_b\right)+\|e\|\cdot e^{\alpha}_a e^{\mu}_b\left(\partial_{\alpha}\delta\omega^{a\hphantom{\mu}b}_{\hphantom{a}\mu}-\partial_{\mu}\delta\omega^{a\hphantom{\alpha}b}_{\hphantom{a}\alpha}\right).
\end{equation}

This is an identity.  Suppose now, for simplicity, that we were making the variation around the pure Weitzenb{\" o}ck frame with $\omega=0$, $\Lambda=I$, and $\delta\omega^{a\hphantom{\mu}b}_{\hphantom{a}\mu}=-\partial_{\mu}\lambda^{ab}$. Then in the left hand side of (\ref{Identity}) we have our variation $\delta_{\omega}S_{{\mathfrak W}^{\prime}}$, while the right hand side shows that it vanishes given that appropriate boundary conditions for $\delta\omega$ are applied.

 Obviously, it should also be true for generic flat connections since, modulo surface terms, the action was locally Lorentz invariant. Explicitly it can be checked by substituting
 \begin{equation} \label{substitute}
T^{\alpha}_{\hphantom{\alpha}\mu\nu}={\mathop T\limits^{\mathfrak W}}{\vphantom{T}}^{\alpha}_{\hphantom{\alpha}\mu\nu}+\omega^{\alpha}_{\hphantom{\alpha}\mu\nu}-\omega^{\alpha}_{\hphantom{\alpha}\nu\mu}
\end{equation}
into the identity (\ref{Identity}) which produces new $\omega\cdot\delta\omega$ terms in the left hand side. And those are precisely the ones which turn the last term in the right hand side into the linear variation of the curvature:
\begin{equation}
\label{Identitycov}
\| e\|\cdot\left(T^{\mu}_{\hphantom{\mu}\alpha\nu}+2T_{\nu}\delta_{\alpha}^{\mu}\right)\delta\omega^{\alpha\hphantom{\mu}\nu}_{\hphantom{\alpha}\mu}=2\partial_{\mu}\left(\|e\|\cdot e^{\alpha}_a\delta\omega^{a\hphantom{\alpha}b}_{\hphantom{a}\alpha}e^{\mu}_b\right)+\|e\|\cdot e^{\alpha}_a e^{\mu}_b\eta^{bc}\left({\mathfrak D}_{\alpha}\delta\omega^{a}_{\hphantom{a}\mu c}-{\mathfrak D}_{\mu}\delta\omega^{a}_{\hphantom{a}\alpha c}\right).
\end{equation}
Relation (\ref{Identitycov}) is a covariant version of identity (\ref{Identity}). Of course, all that we have done is just that we have explicitly checked that
$$\delta_{\omega}\mathbb T=-\delta_{\omega}\left(2 \mathop\bigtriangledown\limits^{(0)}{\vphantom{\bigtriangledown}}_{\mu}T^{\mu}\right)+\delta_{\omega} R(\omega)\,,$$
in accordance with (\ref{telepid}). In the left hand side of (\ref{Identitycov}) we have the variation $\delta_{\omega}S_{{\mathfrak W}^{\prime}}$.
The $\delta_{\omega} R(\omega)$ term vanishes since our variations are still performed in the flat class, and the conclusion is unchanged. Again, in doing the explicit calculation one would obtain from partial derivatives of $\delta\omega$ given by (\ref{inertvar}) the terms on the form $(\Lambda^{-1})(\partial\lambda)(\partial\Lambda)=(\Lambda^{-1})(\partial\lambda)\Lambda\cdot(\Lambda^{-1})(\partial\Lambda)=(\delta\omega)\cdot\omega$ which make compensation for spin connection terms in full $\mathfrak D$-derivatives. Detailed calculations are a bit time consuming, but nevertheless the main idea is quite clear.

To summarise, one may incorporate an inertial connection, parameterised by a Lorentz matrix, into the action without inducing new equations of motions. The variation of the action with respect to inertial spin connection is a surface term and vanishes identically given appropriate boundary conditions. This is a peculiar feature of the teleparallel equivalent of General Relativity. In Section V we will see that, for extended models, the variation does not vanish identically. However, it vanishes if the field equations for tetrads are satisfied.

\subsection{Lagrange multiplier approach}

Equivalently, one could also impose $R^a_{\hphantom{a}b\mu\nu}=0$ with a Lagrange multiplier instead of condition (\ref{inertial}). 
This is of course, the usual and perfectly legitimate way to fix the gauge.  
Then the action would be
\begin{equation}
\label{Lagrtelep}
S_{\mathfrak{LW}}=-\int d^4 x \| e\|\cdot \left({\mathbb T}(e,\omega)+\lambda_a^{\hphantom{a}b\mu\nu}R^a_{\hphantom{a}b\mu\nu}(\omega)\right)
\end{equation}
where $\lambda_a^{\hphantom{a}b\mu\nu}$ is a Lagrange multiplier with the symmetry properties $\lambda^{ab\mu\nu}=-\lambda^{ab\nu\mu}$ and $\lambda^{ab\mu\nu}=-\lambda^{ba\mu\nu}$.

Then the variation of action (\ref{Lagrtelep}) with respect to $\lambda$ yields
$$R^a_{\hphantom{a}b\mu\nu}(\omega)=0$$
which is equivalent to (\ref{inertial}), at least locally. Variation with respect to $e$ results in
\begin{equation*}
\mathop\bigtriangledown\limits^{(0)}{\vphantom{\bigtriangledown}}_{\alpha} S_{\beta}^{\hphantom{\beta}\mu\alpha}-S^{\alpha\mu\nu}\left(T_{\alpha\beta\nu}+K_{\alpha\nu\beta}\right)+\frac12 \left({\mathbb T}+\lambda_a^{\hphantom{a}b\alpha\nu}R^a_{\hphantom{a}b\alpha\nu}\right)\delta^{\mu}_{\beta}=0
\end{equation*}
which is equivalent to (\ref{telepeom}) due to the previous equation. Finally, the variation with respect to $\omega^{a\hphantom{\nu}c}_{\hphantom{a}\nu}$ gives an equation for the Lagrange multiplier:
$${\mathfrak D}_{\mu}\left(\|e\| \lambda_a^{\hphantom{a}b\mu\nu}\right)=0$$
which can be rewritten equivalently as
$$\mathop\bigtriangledown\limits^{(0)}{\vphantom{\bigtriangledown}}_{\mu}\lambda_{\alpha}^{\hphantom{\alpha}\beta\mu\nu}-K^{\gamma}_{\hphantom{\gamma}\mu\alpha}\lambda_{\gamma}^{\hphantom{\gamma}\beta\mu\nu}+K^{\beta}_{\hphantom{\beta}\mu\gamma}\lambda_{\alpha}^{\hphantom{\alpha}\gamma\mu\nu}=0.$$

Therefore we have an equivalent version of covariantised teleparallel gravity together with an equation for the Lagrange multiplier. For this version of teleparallel gravity, see also the book of Blagojevi{\' c} \cite{blagojevic}, and for the implementation in metric-affine gravity, the paper \cite{Obukhov:2002tm}. As the $\lambda_a^{\hphantom{a}b\mu\nu}$ does not enter the equation of motion for the tetrad, we don't need to consider it further, but one may count its degrees of freedom and confirm that the system is consistent. 

\subsection{An $\omega$-blind equivalent to GR}

Finally, we would also like to notice that if we take an action
\begin{equation}
\label{newact}
S_{\omega-free}=-\int d^4 x \| e\|\left( \vphantom{\int}{\mathbb T}- e^{\mu}_a R^a_{\hphantom{\alpha}b\mu\nu}(\omega) e^{\nu}_c \eta^{bc}\right),
\end{equation}
then it is equivalent to the Einstein-Hilbert action modulo a surface term, see (\ref{telepid}) with an arbitrary spin connection, not necessarily inertial. The spin connection makes only a superficial appearance here, and one can formulate equations of general relativity in terms of arbitrary spin connection with geometries containing both curvature and torsion. Unfortunately, a quick look at variations shows that a generalisation with non-linear functions of the Lagrangian density in (\ref{newact}) does not give a non-trivial model. However, it is probably the way which should be taken if one aims at having a fully covariant model with an arbitrary spin connection in a pure tetrad formulation.

\section{Extensions and Discussion} 

As we have already mentioned in the Introduction, it would be interesting to try making modifications of GR in the teleparallel framework. One natural generalisation of GR in the teleparallel formulation is given by $f(T)$ models, another might be formulated with other torsion scalars,
${\tilde T} =  \frac{c_1}4 T_{\alpha\beta\mu}T^{\alpha\beta\mu}+\frac{c_2}2 T_{\alpha\beta\mu}T^{\beta\alpha\mu}-c_3 T_{\mu}T^{\mu}.$
The breakdown of local Lorentz invariance by the Weitzenb{\" o}ck choice of connection is real here, and new modes appear in the pure-tetrad approach. Of course, by taking into account an arbitrary inertial spin connection achieves the covariance of the proper teleparallel gravity theory. However, the new degrees of freedom remain there. For example, in spherically symmetric cases, a nice spherically symmetric prescription for the tetrad might require non-zero spin connection \cite{Krssak:2015rqa}. Of course, in a rotated frame, one will be able to restore the Weitzenb{\" o}ck connection, but at the price of a more involved solution for the tetrad. In other words, the formulation with arbitrary inertial spin connection resolves the issue of "good and bad tetrads" \cite{Tamanini:2012hg}.

The covariantisation procedure works differently in generalised teleparallel gravities. Since the dependence on the spin connection in generalised models cannot be reduced to a surface term, the variation $\delta_{\omega}$ produces non-trivial equations of motion. However, admissible variations of $\omega$ in the inertial class amount to local Lorentz transformations. Note also that a covariantised action is, by definition, identically invariant under simultaneous local Lorentz transformation of the spin connection and the tetrad (and other non-trivially transforming fields if there are some). Therefore, the stationarity of the action under local Lorentz transformations of the spin connection is equivalent to that under local Lorentz rotations of tetrads. The latter is already ensured given that the equations of motion for the tetrad are satisfied since the local Lorentz rotation is nothing but a special class of variations of the tetrad. We conclude that the field equations for the spin connection, though non-trivial, are redundant with that of the tetrad. We would like to illustrate this very important point in the case of $f(T)$ gravity. In this Section we discuss it at the level of variations, and in the next Section we show how it works with equations of motion.

Let us consider an $f(T)$ model with inertial spin connection,
\begin{equation}
\label{covfT}
S_{f(T)}=-\int d^4 x \| e\|\cdot f\left({\mathbb T}(e,\omega(\Lambda)\right).
\end{equation}
Variation with respect to $\omega$ is easily performed using (\ref{telepid}) and (\ref{omtorvecvar}), and yields
\begin{equation}
\label{ftomvar}
\delta_{\omega} S_{f(T)}=-\int d^4 x \|e\| \left(\mathop\bigtriangledown\limits^{(0)}{\vphantom{\bigtriangledown}}_{\mu} f^{\prime}({\mathbb T})\right) e^{\nu}_a e^{\mu}_b\eta^{cb}\delta\omega^{a}_{\hphantom{a}\nu c}
\end{equation}
which gives non-trivial equations of motion even for purely inertial spin connections and their variations given by (\ref{inertial}) and (\ref{inertvar}).

However, the resulting equation is already contained in the equations which are given by variations with respect to tetrads. Indeed, among the variations of tetrads, one option is $e^a_{\mu}\to \Lambda^a_b e^b_{\mu}$. This is a local Lorentz rotation, and using (\ref{telepid}) we have for this type of variation $f^{\prime} \cdot\delta{\mathbb T}=-2f^{\prime}\cdot \mathop\bigtriangledown\limits^{(0)}{\vphantom{\bigtriangledown}}_{\mu}\delta T^{\mu}$. One can see that the variation of the torsion vector will take the same form as $e^{\nu}_a e^{\mu}_b\eta^{cb}\delta\omega^{a}_{\hphantom{a}\nu c}$ with $\delta\omega$ being the variation of the (inertial) spin connection under the same Lorentz transformation. For example, around the Weitzenb{\" o}ck frame the variation of the spin connection will take the form of $\delta\omega^{a}_{\hphantom{a}\alpha b}=-\partial_{\alpha}\lambda^a_b$ and $\delta T_{\mu}=e^{\alpha}_a (\partial_{\alpha}\lambda^a_b)e^b_{\mu}$. At the same time, an infinitesimal Lorentz rotation of tetrads reads $\delta e^a_{\mu}=\lambda^a_b e^b_{\mu}$. In the torsion vector $\mathop T\limits^{\mathfrak W}{\vphantom{T}}_{\mu}= e^{\alpha}_a \partial_{\mu}e^a_{\alpha}-e^{\alpha}_a \partial_{\alpha}e^a_{\mu}$, the first term is locally Lorentz invariant (and equal to $\frac{1}{\|e\|}\partial_{\mu}\|e\|$), while the variation of the second one is $-e^{\alpha}_a (\partial_{\alpha}\lambda^a_b)e^b_{\mu}$. Therefore, the variation with respect to a purely inertial connection is the same as one gets by a special variation of the tetrads.

Let us discuss it also without the trick of using the identity (\ref{telepid}). Varying the connection we have
$$\delta_{\omega} S_{f(T)}=-\int d^4 x \|e\| f^{\prime}({\mathbb T}) S_{\alpha}^{\hphantom{\alpha}\mu\nu}\delta T^{\alpha}_{\hphantom{\alpha}\mu\nu}=-2\int d^4 x \|e\| f^{\prime}({\mathbb T}) S_{\alpha}^{\hphantom{\alpha}\mu\nu}e^{\alpha}_a e^{b}_{\nu}\delta \omega^{a}_{\hphantom{a}\mu b}$$
Using (\ref{inertvar}), we finally have the variation
$$\delta_{\omega} S_{f(T)}=2\int d^4 x \|e\| f^{\prime}({\mathbb T}) S_{\alpha}^{\hphantom{\alpha}\mu\nu}e^{\alpha}_a (\Lambda^{-1})^a_c (\partial_{\mu}\lambda^c_d)\Lambda^d_b e^{b}_{\nu}$$
which must vanish for arbitrary $\lambda^c_d\in so(1,3)$.

Now we look at the variations with respect to the tetrads, $\delta_e S_{f(T)}=-\int d^4 x\left(f({\mathbb T})\delta\|e\|+\|e\|  f^{\prime}({\mathbb T})\delta {\mathbb T} \right)$. The $\delta S$ must vanish for all possible variations of tetrads. This requirement yields the equations of motion. In particular, it must vanish for variations of the form $e\to (\exp{\tilde\lambda})\cdot e$, or infinitesimally $\delta e^a_{\mu}=\tilde\lambda^a_b e^b_{\mu}$. In such a case, we have $\delta\|e\|=0$, and $\delta_e {\mathbb T}=S_{\alpha}^{\hphantom{\alpha}\mu\nu}\delta_e T^{\alpha}_{\hphantom{\alpha}\mu\nu}$ since $\delta_e g_{\mu\nu}=0$. Taking then into account that in our case $\delta_e T^{\alpha}_{\hphantom{\alpha}\mu\nu}=e^{\alpha}_a\left(({\mathfrak D}_{\mu}{\tilde\lambda}^a_b)e^b_{\nu}-({\mathfrak D}_{\nu}{\tilde\lambda}^a_b)e^b_{\mu}\right)$, we have
$$\left.\delta_{e} S_{f(T)}\vphantom{\int}\right|_{\delta e=\tilde\lambda e}=2\int d^4 x \|e\| f^{\prime}({\mathbb T}) S_{\alpha}^{\hphantom{\alpha}\mu\nu}e^{\alpha}_a({\mathfrak D}_{\mu}{\tilde\lambda}^a_b)e^b_{\nu}.$$
As was shown after the formula (\ref{inertvar}), if we take ${\tilde\lambda}^a_b=-(\Lambda^{-1})^a_c\lambda^c_d\Lambda^d_b$ then
$${\mathfrak D}_{\mu}{\tilde\lambda}^a_b=-(\Lambda^{-1})^a_c (\partial_{\mu}\lambda^c_d)\Lambda^d_b.$$
Now we see that $\left.\delta_{e} S_{f(T)}\vphantom{\int}\right|_{\delta e=\tilde\lambda e}=-\delta_{\omega} S_{f(T)}$. Equations of motion which follow from $\delta_{\omega}$ in the inertial class are already there as a special case of those equations which can be derived by variations of tetrads. This is how it must be since if one makes a local Lorentz rotation self-consistently changing both $e$ and $\omega$, then the covariantised action is invariant which means that the two separate variations do cancel each other.

Let us stress again that it is generic for any models of covariantised (by inertial spin connection) gravities of teleparallel type. Making the variation $\Lambda^a_b \to (\exp(\lambda))^a_c \Lambda^c_b$ in the inertial connection (\ref{inertial}) results in precisely the same change in action, though with the opposite sign, as $e^a_{\mu}\to (\exp(\tilde\lambda))^a_c e^c_{\mu}$ with an appropriate $\tilde\lambda$. This is the bare essence of covariantisation. Making two variations simultaneously implies no change at all if all the derivatives were Lorentz-covariant. Of course, the same conclusion can be derived with the Lagrange multiplier approach. 

With usual gauge theories in mind, it might seem strange that the connection did not lead to any new non-trivial equations of motion. Of course, this reflects the freedom of choosing a frame, and the striking difference from the usual gauge theories stems from the requirement of the purely inertial spin connection. In an electromagnetic analogy, it would amount to gauging the global $U(1)$ of Dirac fermions by a pure gauge field. In other words, the covariant derivatives would feature a gradient of a scalar $\partial_{\mu}-ie\partial_{\mu}\phi$ instead of the vector potential $\partial_{\mu}-ieA_{\mu}$. It is nevertheless covariant under local transformations (when $\phi$ is properly shifted simultaneously with the local phase rotation of the spinor) but it is not what we usually call gauging. Obviously, the variation with respect to $\phi$ gives the conservation of the fermion current which was already there to begin with, in the Dirac equation. This would be the analogy to our variation with respect to the purely inertial spin connection. The analogy of the Lagrange multiplier method would be to vary with respect to a general $A_{\mu}$, but impose the vanishing of the electromagnetic curvature $F=0$; and trivially, the analogy of the pure-tetrad Weitzenb\"ock approach would to set $A_{\mu}=0$.  
One has to bare in mind that the local Lorentz invariance in teleparallel gravity conceptually very different from the local $U(1)$ invariance in electrodynamics.

Finally, we remark that in gauge gravity, the problem of covariantisation does not end but only properly begins by restoring the Lorentz invariance. 
To consider invariance under the full Poincar\'e group, or a larger symmetry group, one first notices that the tetrad $e^a_\mu$ cannot be quite the same as a translation gauge potential $t^a_\mu$, as the former must transform like a tensor, the latter like a connection. However, one may introduce a covariantly transforming vector $x^a$, to define the relation between the two as
$e^a_\mu = \partial_\mu x^a +  t^a_\mu + \omega^a_{\phantom{a}\mu b}x^b$. One then obtains the desired projector object, the tetrad, in a consistent, gauge-invariant manner, and by setting $x^a=0$, one recovers the same result. This is basically the tacit choice one is making if considering teleparallel gravity as a translation gauge theory. But how to justify the choice $x^a=0$ and what is this vector $x^a$? Let us only mention that it is our pedestrian-formalism equivalent of the most interesting objects - from development operators of Cartan geometry through tensorial left-overs of nonlinearly realised connections to the cartwheels of the idealised waywiser - and refer the reader, for the state of art in the geometric formulation of spontaneously broken first order gauge gravity, to Refs. \cite{Westman:2013mf,Zlosnik:2016fit}.

\section{Equations of motion in $f(T)$ gravity}

In the previous Section we explained that for an extended teleparallel model, i.e. the corresponding ''pure-tetrad'' model covariantised by inertial spin connection, the equations of motion for $\omega$ are either redundant to or equivalent with those for the tetrads. Now we will show how it can be checked directly at the level of equations of motion in the simple but very important case of $f(T)$ gravity. We consider an $f(T)$ model with inertial spin connection,
\begin{equation}
S_{f(T)}=-\int d^4 x \| e\|\cdot f\left({\mathbb T}(e,\omega(\Lambda)\right),
\end{equation}
and want to derive equations of motion.

First we look at the variations with respect to tetrads. We have
$$\delta S_e=-\int d^4 x \|e\|\cdot\left(\vphantom{\int}f(\mathbb T)e^{\mu}_a \delta e^a_{\mu} + f^{\prime}(\mathbb T)\delta\mathbb T\right).$$
For inertial spin connections, the identity (\ref{telepid}) reads $\mathbb T=-\mathop R\limits^{(0)}-2{\mathop\bigtriangledown\limits^{(0)}}_{\mu}T^{\mu}$, and
$$\delta S_e=-\int d^4 x \|e\|\cdot\left(\vphantom{\int}f(\mathbb T)e^{\mu}_a \delta e^a_{\mu} - f^{\prime}(\mathbb T)\delta\mathop R\limits^{(0)}+2\left(\partial_{\mu} f^{\prime}(\mathbb T)\right)\delta(g^{\mu\nu}T_{\nu}) \right).$$
Given that $\delta g_{\mu\nu}=2\eta_{ab}e^a_{\mu}\delta e^b_{\nu}$ and $\delta g^{\mu\nu}=-\left(g^{\mu\rho}e^{\nu}_a+g^{\nu\rho}e^{\mu}_a\right)\delta e^a_{\rho}$, we get
\begin{multline*}
\delta S_e=-\int d^4 x \|e\|\cdot\left(\vphantom{\int}f e^{\mu}_a \delta e^a_{\mu} +2\left(\vphantom{\int}\left({\mathop R\limits^{(0)}}{\vphantom{R}}^{\mu\nu}-{\mathop\bigtriangledown\limits^{(0)}}{\vphantom{R}}^{\mu}{\mathop\bigtriangledown\limits^{(0)}}{\vphantom{R}}^{\nu}+g^{\mu\nu}\mathop\square\limits^{(0)}\right)f^{\prime}\right)\eta_{ab}e^a_{\mu}\delta e^b_{\nu}\right.\\
-\left.2\left(\vphantom{\int}\left(\partial^{\mu} f^{\prime}\right)T_{\nu}+\left(\partial_{\nu} f^{\prime}\right)T^{\mu}\right)e^{\nu}_a\delta e^a_{\mu}+2\left(\partial^{\mu} f^{\prime}\right)\delta T_{\mu} \right).
\end{multline*}

Now we need the variation of the torsion (\ref{etorvar})
$$\delta_e T^{\alpha}_{\hphantom{\alpha}\mu\nu} =  -e^{\alpha}_a T^{\beta}_{\hphantom{\beta}\mu\nu}\delta e^a_{\beta}+e^{\alpha}_a \left({\mathop\mathfrak D}_{\mu}\delta e^a_{\nu}-{\mathop\mathfrak D}_{\nu}\delta e^a_{\mu}\right).$$
For later convenience, note that the connection coefficients from the first term on the left hand side make the Lorentz-covariant derivatives in the second term into the full derivatives of $\delta e$. Given that full derivatives of tetrads vanish, we have
\begin{equation}
\label{betteretorvar}
\delta_e T^{\alpha}_{\hphantom{\alpha}\mu\nu} = \bigtriangledown_{\mu}\left(e^{\alpha}_a \delta e^a_{\nu}\right)-\bigtriangledown_{\nu}\left(e^{\alpha}_a \delta e^a_{\mu}\right),
\end{equation}
and in particular
$$\delta_e T_{\mu} = \partial_{\mu}\left(e^{\alpha}_a \delta e^a_{\alpha}\right)-\bigtriangledown_{\alpha}\left(e^{\alpha}_a \delta e^a_{\mu}\right),$$
Accounting for the difference between the actual connection and the Levi-Civita one, we have
$$\delta_e T_{\mu} = \partial_{\mu}\left(e^{\alpha}_a \delta e^a_{\alpha}\right)-{\mathop\bigtriangledown\limits^{(0)}}_{\alpha}\left(e^{\alpha}_a \delta e^a_{\mu}\right)-K^{\alpha}_{\hphantom{\alpha}\alpha\nu}e^{\nu}_a \delta e^a_{\mu}+K^{\nu}_{\hphantom{\nu}\alpha\mu}e^{\alpha}_a \delta e^a_{\nu}.$$

Now the equations of motion can be obtained straightforwardly. They read:
\begin{equation}
\label{fteom}
f^{\prime}(\mathbb T)\cdot{\mathop R\limits^{(0)}}{\vphantom{R}}^{\mu\nu}+K^{\mu\nu\alpha}\partial_{\alpha} f^{\prime}(\mathbb T)-T^{\mu}\partial^{\nu} f^{\prime}(\mathbb T)+\frac12 f(\mathbb T)\cdot g^{\mu\nu}=0
\end{equation}
where we have used $K^{\alpha}_{\hphantom{\alpha}\alpha\nu}=-T_{\nu}$. Note that the higher derivative terms, typical for $f(R)$ gravity, have cancelled each other as it should be.

Unlike the TEGR case, equation (\ref{fteom}) has a non-trivial antisymmetric part:
\begin{equation}
\label{antisym}
T^{\alpha\mu\nu}\partial_{\alpha} f^{\prime}(\mathbb T)+T^{\nu}\partial^{\mu} f^{\prime}(\mathbb T)-T^{\mu}\partial^{\nu} f^{\prime}(\mathbb T)=0
\end{equation}
where we have used $K^{\mu\nu\alpha}-K^{\nu\mu\alpha}=T^{\alpha\mu\nu}$. We should now show that this equation coincides precisely with the one which comes from the variation of the inertial spin connection.

Indeed, we have the variation of the action with respect to inertial spin connection (\ref{ftomvar}) with $\delta\omega^{a\hphantom{\nu}b}_{\hphantom{a}\nu}={\mathfrak D}_{\nu}{\tilde\lambda}^{ab}$ for arbitrary antisymmetric $\tilde\lambda$:
$$\delta_{\omega} S_{f(T)}=-\int d^4 x \|e\| \left(\partial_{\mu} f^{\prime}({\mathbb T})\right) e^{\mu}_a e^{\nu}_b {\mathfrak D}_{\nu}{\tilde\lambda}^{ab}.$$
Integrating by parts, we have
$$\delta_{\omega} S=\int d^4 x \left(\partial_{\mu} f^{\prime}\right)\cdot \left(\vphantom{\int}\partial_{\nu}\left(\|e\|e^{\mu}_a e^{\nu}_b\right)-\|e\|\left(\omega^{c}_{\hphantom{c}\nu a}e^{\mu}_c e^{\nu}_b+\omega^{c}_{\hphantom{c}\nu b}e^{\mu}_a e^{\nu}_c\right)\right){\tilde\lambda}^{ab}.$$
Using $\partial_{\nu}\|e\|=\|e\|{\mathop\Gamma\limits^{(0)}}{\vphantom{\Gamma}}^{\alpha}_{\nu\alpha}$ and vanishing of the full derivative of the tetrad we get:
$$\delta_{\omega} S=-\int d^4 x \|e\|\left(\partial_{\mu} f^{\prime}\right)\cdot\left(K^{\alpha}_{\hphantom{\alpha}\alpha\beta}e^{\mu}_a e^{\beta}_b+\Gamma^{\mu}_{\nu \beta}e^{\beta}_a e^{\nu}_b\right){\tilde\lambda}^{ab}.$$
Due to antisymmetry of $\tilde\lambda$ it reproduces equation (\ref{antisym}).

\section{Conclusions}

As teleparallel models have become, and rightly so, a matter of interest for building non-standard cosmological models, 
it is very important to thoroughly understand their dynamics. In this paper we have clarified some subtle issues concerning covariatisations of the ''pure-tetrad'' versions
 of these models. We have seen that there are several methods of approaching teleparallel gravity covariantly. They can be classified according to how the variation of the spin connection is performed. 

{\bf 0.} "Weitzenb{\" o}ck variation": $\omega=0$. This is the historical formulation of teleparallel gravity, and it is not covariant.

{\bf 1.} "Fixed omega variation": $\omega$ is totally fixed but arbitrary. It can be thought of as writing the Weitzenb{\" o}ck action in an arbitary frame by substituting the zero connection by a proper quantity of the form of (\ref{inertial}) with a fixed $\Lambda(x)$. Under any fixed choice, the Lagrangian is not invariant, however there is the freedom of making this choice.

{\bf 2.} "Independent variation" of the unrestricted $\omega$, see (\ref{covart}). It results in no gravity at all, $T=0$.

{\bf 3.} "Inertial variation": $\omega$ is varied in the class of inertial spin connections, see (\ref{inertial}) and (\ref{inertvar}). Independent variables are tetrads and Lorentz matrices which parametrise the spin connection. This is a proper covariantisation for both teleparallel equivalent of GR and modified teleparallel models.

{\bf 4.} "Constrained variation": $\omega$ is by itself an independent variable, however its curvature tensor is set to zero with a Lagrange multiplier (\ref{Lagrtelep}). This is equivalent to previous option plus an equation for the Lagrange multiplier.

{\bf 5.} "Compensating variation": decoupled $\omega$, see (\ref{newact}). Strictly speaking, it is no longer a teleparallel model. The spin connection is arbitrary indeed, and gravity can be expressed as a combination of torsion and curvature, in any proportions we like. It works for the various teleparallel equivalents of GR thus introduced. Whether it can be used for modified gravity scenarios, remains to be seen.
\newline
\newline
{\bf Acknowledgements.}
We thank Yen Chin Ong, Martin Kr{\v s}{\v s}{\' a}k, Nicola Tamanini and the anonymous referees for useful comments on the manuscript.
AG was supported by the Dynasty Foundation grant and by Russian Science Foundation (project 16-11-10218). AG is also grateful to Saint Petersburg State University for the travel grant 11.42.1061.2016; and to Universities of Helsinki and Jyv{\" a}skyl{\" a}, and to NORDITA for hospitality and partial support during his visits.

\bibliography{covartelepar}

\begin{thebibliography}{10}

\bibitem{Clifton:2011jh}
T.~Clifton, P.~G. Ferreira, A.~Padilla, and C.~Skordis, ``{Modified Gravity and
  Cosmology},'' {\em Phys. Rept.}, vol.~513, pp.~1--189, 2012.

\bibitem{Goenner2004}
H.~F.~M. Goenner, ``On the history of unified field theories,'' {\em Living
  Reviews in Relativity}, vol.~7, no.~1, p.~2, 2004.

\bibitem{moller}
C.~M{\o}ller, {\em Conservation Law and Absolute Parallelism in General
  Relativity}.
\newblock Matematisk-fysiske skrifter K. Danske videnskabernes selskab 31
  (1959), no. 14.

\bibitem{newtensor}
V.~C. de~Andrade, L.~C.~T. Guillen, and J.~G. Pereira, ``{Gravitational energy
  momentum density in teleparallel gravity},'' {\em Phys. Rev. Lett.}, vol.~84,
  pp.~4533--4536, 2000.

\bibitem{blagojevic}
M.~Blagojevic, {\em {Gravitation and gauge symmetries \rm IoP, Bristol }}.
\newblock 2002.

\bibitem{Aldrovandi:2013wha}
R.~Aldrovandi and J.~G. Pereira, {\em {Teleparallel Gravity}}, vol.~173 of {\em
  Fundamental Theories of Physics}.
\newblock Dordrecht: Springer, 2013.

\bibitem{Linder:2010py}
E.~V. Linder, ``{Einstein's Other Gravity and the Acceleration of the
  Universe},'' {\em Phys. Rev.}, vol.~D81, p.~127301, 2010.
\newblock [Erratum: Phys. Rev.D82,109902(2010)].

\bibitem{LSB1}
B.~Li, T.~P. Sotiriou, and J.~D. Barrow, ``{$f(T)$ gravity and local Lorentz
  invariance},'' {\em Phys. Rev.}, vol.~D83, p.~064035, 2011.

\bibitem{LSB2}
T.~P. Sotiriou, B.~Li, and J.~D. Barrow, ``{Generalizations of teleparallel
  gravity and local Lorentz symmetry},'' {\em Phys. Rev.}, vol.~D83, p.~104030,
  2011.

\bibitem{LSB3}
B.~Li, T.~P. Sotiriou, and J.~D. Barrow, ``{Large-scale Structure in f(T)
  Gravity},'' {\em Phys. Rev.}, vol.~D83, p.~104017, 2011.

\bibitem{Ferraro:2011us}
R.~Ferraro and F.~Fiorini, ``{Non trivial frames for f(T) theories of gravity
  and beyond},'' {\em Phys. Lett.}, vol.~B702, pp.~75--80, 2011.

\bibitem{Tamanini:2012hg}
N.~Tamanini and C.~G. Boehmer, ``{Good and bad tetrads in f(T) gravity},'' {\em
  Phys. Rev.}, vol.~D86, p.~044009, 2012.

\bibitem{Ong:2013qja}
Y.~C. Ong, K.~Izumi, J.~M. Nester, and P.~Chen, ``{Problems with Propagation
  and Time Evolution in f(T) Gravity},'' {\em Phys. Rev.}, vol.~D88, p.~024019,
  2013.

\bibitem{Chen:2014qtl}
P.~Chen, K.~Izumi, J.~M. Nester, and Y.~C. Ong, ``{Remnant Symmetry,
  Propagation and Evolution in $f$(T) Gravity},'' {\em Phys. Rev.}, vol.~D91,
  no.~6, p.~064003, 2015.

\bibitem{kop}
W.~Kopczynski, ``Problems with metric-teleparallel theories of gravitation,''
  {\em Journal of Physics A: Mathematical and General}, vol.~15, no.~2, p.~493,
  1982.

\bibitem{Blagojevic:2013xpa}
M.~Blagojevi{\'c} and F.~W. Hehl, eds., {\em {Gauge Theories of Gravitation}}.
\newblock Singapore: World Scientific, 2013.

\bibitem{f(T)review}
Y.-F. Cai, S.~Capozziello, M.~De~Laurentis, and E.~N. Saridakis, ``{f(T)
  teleparallel gravity and cosmology},'' {\em Rept. Prog. Phys.}, vol.~79,
  no.~10, p.~106901, 2016.

\bibitem{Krssak:2015rqa}
M.~Kr{\v s}{\v s}{\' a}k and J.~G. Pereira, ``{Spin Connection and
  Renormalization of Teleparallel Action},'' {\em Eur. Phys. J.}, vol.~C75,
  no.~11, p.~519, 2015.

\bibitem{Krssak:2015lba}
M.~Kr{\v s}{\v s}{\' a}k, ``{Holographic Renormalization in Teleparallel
  Gravity},'' {\em Eur. Phys. J.}, vol.~C77, no.~1, p.~44, 2017.

\bibitem{Krssak}
M.~Kr{\v s}{\v s}{\'a}k and E.~N. Saridakis, ``{The covariant formulation of
  f(T) gravity},'' {\em Class. Quant. Grav.}, vol.~33, no.~11, p.~115009, 2016.

\bibitem{NesterYo}
J.~M. Nester and H.-J. Yo, ``{Symmetric teleparallel general relativity},''
  {\em Chin. J. Phys.}, vol.~37, p.~113, 1999.

\bibitem{Adak:2005cd}
M.~Adak, M.~Kalay, and O.~Sert, ``{Lagrange formulation of the symmetric
  teleparallel gravity},'' {\em Int. J. Mod. Phys.}, vol.~D15, pp.~619--634,
  2006.

\bibitem{Bahamonde:2015hza}
S.~Bahamonde and M.~Wright, ``{Teleparallel quintessence with a nonminimal
  coupling to a boundary term},'' {\em Phys. Rev.}, vol.~D92, no.~8, p.~084034,
  2015.

\bibitem{Obukhov:2002tm}
{\relax Yu}.~N. Obukhov and J.~G. Pereira, ``{Metric affine approach to
  teleparallel gravity},'' {\em Phys. Rev.}, vol.~D67, p.~044016, 2003.

\bibitem{Westman:2013mf}
H.~F. Westman and T.~G. Zlosnik, ``{Exploring Cartan gravity with dynamical
  symmetry breaking},'' {\em Class. Quant. Grav.}, vol.~31, p.~095004, 2014.

\bibitem{Zlosnik:2016fit}
T.~G. Zlosnik and H.~F. Westman, ``{A first-order approach to conformal
  gravity},'' arXiv:1601.00567.

\end{thebibliography}

\end{document}